\documentclass[fontsize=11pt,parskip=full,dvipsnames]{article}
\usepackage{amsmath,amssymb,amsfonts,amsthm,mathabx}
\usepackage[dvips]{graphics}
\usepackage{enumerate}
\usepackage{mathrsfs}
\usepackage{subfigure}
\usepackage{color}
\usepackage{setspace}  %
\usepackage{hyperref}
\usepackage{float}
\usepackage{microtype} %
\usepackage{xfrac} %

\usepackage[font=small,skip=5pt]{caption} %

\hypersetup{hidelinks = true} %

\usepackage[T1]{fontenc}

\usepackage{authblk} %
\usepackage{bbm} %

\usepackage{enumerate} %

\usepackage[backend=biber, style=nature,url=false,isbn=false,uniquelist=false]{biblatex}
\AtEveryBibitem{\clearfield{month}}
\AtEveryBibitem{\clearfield{day}}
\AtEveryBibitem{\clearfield{language}}

\DeclareSymbolFont{AMSb}{U}{msb}{m}{n}

\usepackage{geometry}
\usepackage{color,soul}

\usepackage{amsmath}
\DeclareMathOperator*{\argmax}{argmax} 
\DeclareMathOperator*{\Tr}{Tr} 
 
\DeclareMathOperator*{\diag}{diag}

\title{PyGenStability: Multiscale community detection with generalized Markov Stability}

\author[1,*]{Alexis Arnaudon}
\author[1,*]{Juni Schindler}
\author[2]{Robert L. Peach}
\author[3]{Adam Gosztolai}
\author[4]{Maxwell Hodges}
\author[5]{Michael T. Schaub}
\author[1]{Mauricio Barahona}
\affil[1]{\normalsize{Department of Mathematics, Imperial College London, London, UK}}
\affil[2]{\normalsize{Department of Neurology, University Hospital Würzburg, Würzburg, Germany}}
\affil[3]{\normalsize{Signal Processing Laboratory (LTS2), EPFL, Lausanne, Switzerland}}
\affil[4]{\normalsize{Spotify, London, UK}}
\affil[*]{These two authors contributed equally to this paper.}

\date{}

\begin{document}\maketitle

\begin{abstract}\noindent
    We present PyGenStability, a general-use Python software package that provides a suite of analysis and visualisation tools for unsupervised multiscale community detection in graphs. PyGenStability finds optimized partitions of a graph at different levels of resolution by maximizing the generalized Markov Stability quality function with the Louvain or Leiden algorithms. The package includes automatic detection of robust graph partitions and allows the flexibility to choose quality functions for weighted undirected, directed and signed graphs, and to include other user-defined quality functions. 
\end{abstract}


\section{Introduction}

Unsupervised community detection, or graph clustering, can be traced back to early work in social network analysis in the 1950s~\cite{schindlerCommunityVagueOperator2023} and has become a fundamental data analysis tool in the physical and life sciences, as well as in quantitative social science~\cite{fortunatoCommunityDetectionGraphs2010}. 
Various notions of communities (with associated algorithms) have been developed stemming from different mathematical concepts~\cite{schaubManyFacetsCommunity2017}, including normalized cut~\cite{shi}, non-negative matrix factorisation~\cite{Du}, or modularity maximisation~\cite{girvanCommunityStructureSocial2002}, among others. Furthermore, in many cases of theoretical and practical interest, graphs have relevant structure at multiple scales (or levels of resolution)~\cite{schaubMarkovDynamicsZooming2012}; hence extensions that can deal with multiscale graphs have been proposed based on, e.g., the dynamics of random walks and diffusion processes on graphs~\cite{delvenneStabilityGraphCommunities2010,Lambiotte2008arXiv0812},
graph signal processing~\cite{Tremblay}, or discrete geometry~\cite{Gosztolai2021}. Indeed, recent work has emphasized that, as for other problems in data clustering, a universally best algorithm for community detection cannot exist~\cite{peelGroundTruthMetadata2017}, and that different partitions may thus be needed to describe various aspects of the structure of a graph~\cite{schaubManyFacetsCommunity2017}.

In this spirit, we introduce \texttt{PyGenStability}, a publicly available software package for multiscale community detection based on the optimization of the generalized Markov Stability (MS) multiscale quality function. The MS framework, which was developed in a series of papers ~\cite{delvenneStabilityGraphCommunities2010,delvenneStabilityGraphPartition2013,Lambiotte2008arXiv0812,lambiotteRandomWalksMarkov2014, schaubMarkovDynamicsZooming2012,schaub2019multiscale}, exploits graph diffusion processes to uncover graph partitions at different levels of resolution and has the flexibility to accommodate different notions of graph communities through the modification of a quality function. However, MS has been missing efficient software to boost its adoption by practitioners in data science and in different academic domains. \texttt{PyGenStability} fills this gap and provides a versatile Python package that encompasses several useful variants of the generalized MS quality function to allow for the analysis of undirected, directed, and signed graphs, as well as including fast approximations for large graphs.

The multiscale community detection problem is defined as the following optimization problem.
Given a graph $\mathcal{G}$ with $N$ vertices,
\texttt{PyGenStability} finds a series of optimized graph partitions at different values of a scale parameter $t$ by maximizing the \textit{generalized Markov Stability} function~\cite{schaub2019multiscale}:
\begin{align}
        H^*(t) = \argmax_H \, Q_{gen}(t,H) :=  \argmax_{H} \, \Tr \left [H^T\left ( F(t) - \sum_{k=1}^m v_{2k-1} v_{2k}^T\right)H\right ]\, , 
    \label{eqn:gen_mod}
\end{align}
where the output is a series of $N \times c$  indicator matrices $H^*(t)$ describing the optimised (hard) partitions of the $N$ nodes into $c$ communities for different values of the scale parameter $t$.
Here $F(t)$ is an $N \times N$ \textit{node similarity matrix} that measures the similarity between the nodes of the graph as a function of $t$, and $\{v_k\}_{k=1}^{2m}$ is a set of $N$ dimensional node vector pairs that encode a \textit{null model} of rank $m$. The null model provides the reference against which the quality of the partition is compared. The scale parameter $t$, sometimes referred to as the \textit{Markov time} or \textit{Markov scale}, regulates the coarseness of the partition $H^*(t)$, and the optimization is solved \textit{across all scales}, i.e., for a range of values $t > 0$ that spans from the finest to the coarsest resolution. 

Our package implements constructors to design various quality matrices $F(t)$ and null models $\{v_k\}$, each of which yields different notions of quality and balance for the graph partitions and ensuing communities.
For example, we can use the graph heat kernel $F(t) = \Pi\exp(-Lt)$ with a null model of rank $m=1$ defined by $v_1=v_2 = \pi$, where $L$ is a graph Laplacian, the vector $\pi$ is the stationary distribution of the associated Markov process, and $\Pi=\diag(\pi)$ \cite{delvenneStabilityGraphPartition2013,lambiotteRandomWalksMarkov2014}. In this case, $F(t) = \Pi \exp(-Lt)$ corresponds to the transition probabilities of a Markov process over time $t>0$, and Eq.~\eqref{eqn:gen_mod} can be viewed as optimizing the partition of the graph into subgraphs where the Markov process is more likely to remain contained over time $t$, as compared to the expected behaviour at stationarity. 
Therefore this dynamic viewpoint allows for scanning across different levels of coarseness through $t$. We describe below in \ref{sec:constructors} the implementation of various constructors for different graph types such as weighted, directed, and even signed graphs.

To carry out the combinatorial optimization of the generalized MS quality function of Eq.~\eqref{eqn:gen_mod}, \texttt{PyGenStability} provides a Python wrapper around the C++ implementation of two efficient greedy algorithms: the Louvain~\cite{blondelFastUnfoldingCommunities2008} and Leiden~\cite{traagLouvainLeidenGuaranteeing2019} optimizers.
Further, it is easy to implement other graph clustering algorithms that can be written as a maximization of a generalized function $Q_{gen}(t,H)$~\cite{schaub2019multiscale}, making the package easily extendable. \texttt{PyGenStability} also includes a suite of analysis and visualization tools to process and analyse multi-scale graph partitions, and to facilitate the automatic detection of robust partitions at different scales~\cite{schindlerMultiscaleMobilityPatterns2023}. 
As its output, \texttt{PyGenStability} provides a description of the graph in terms of a sequence of robust partitions $H^*(t_i)$ at scales $t_i$ of increasing coarseness, yet not necessarily hierarchical.

\section{Implementation}

\subsection{Overall organization}
\label{sec:organization}
The Python package \texttt{PyGenStability} consists of four parts:
\begin{enumerate}
    \item {\it Quality function and null model constructors:} This module inputs the node similarity matrix function $F(t)$ and a null model described by vectors $\{v_k\}$. To maintain the flexibility of the package, we provide an object-oriented module to write user-defined constructors for these objects. 
    To facilitate usage, we also provide several constructors already implemented that can be chosen by the user (see Section~\ref{sec:constructors}).
    \item {\it Generalized Markov Stability maximizers:} The combinatorial optimization of the generalized Markov Stability function (Eq.~\ref{eqn:gen_mod}) is carried out by interfacing with two fast algorithms: (i) Louvain~\cite{blondelFastUnfoldingCommunities2008} or (ii) Leiden~\cite{traagLouvainLeidenGuaranteeing2019}, both implemented in C++. The choice of the optimizer is left to the user: Louvain is widely and successfully used in many fields; Leiden is a recent refinement of Louvain, which introduces several improvements, e.g., ensuring connected communities. 
    \item {\it Post-processing tools:} We provide several steps to facilitate the detection of robust optimized partitions, and to ease the analysis of multiscale clusterings (see Section~\ref{sec:post-processing}).
    \item {\it Plotting:} We provide a module to plot the multiscale clustering results, as illustrated in Fig.~\ref{fig:example}.
\end{enumerate}
These four components are tied together via a single, configurable entry point, or can be used independently, depending on user needs. 

\subsection{Quality function constructors}
\label{sec:constructors}
To aid users, we have already implemented several constructors for different versions of the generalized Markov Stability quality function based on graph Laplacians.
For weighted, undirected graphs we have included~\cite{delvenneStabilityGraphCommunities2010,delvenneStabilityGraphPartition2013,lambiotteRandomWalksMarkov2014}:
    (i) MS based on the continuous-time random-walk (normalized) graph Laplacian; 
    (ii) MS based on the continuous-time combinatorial graph Laplacian;
    (iii) linearized MS based on the normalized graph Laplacian (also referred to as `modularity with resolution parameter') 
    for a more computationally efficient analysis of larger graphs (see Fig.~\ref{fig:benchmark}).
 For weighted, directed graphs we have included~\cite{lambiotteRandomWalksMarkov2014,schaub2019multiscale}: 
    (iv) MS based on the continuous-time random-walk Laplacian with teleportation;
    (v) linearized MS for the random-walk Laplacian with teleportation (more efficient for larger graphs).
For weighted, signed graphs:
    (vi) MS based on the signed Laplacian as given in~\cite{schaub2019multiscale};
    (vii) a version of signed modularity with resolution~\cite{gomezAnalysisCommunityStructure2009} 
    (more efficient for larger graphs).
More detailed information about these constructors can be found in our code documentation hosted on GitHub. Our object-oriented module facilitates the simple implementation of further custom constructors. 

\subsection{Post-processing tools}
\label{sec:post-processing}

\paragraph{Quantifying the robustness of partitions through the Normalized Variation of Information.}

Louvain and Leiden are both greedy algorithms, which provide local maxima to the combinatorial optimization problem~\eqref{eqn:gen_mod} without guarantees of global optimality. 
The optimization can thus produce different maxima depending on the starting point of the iterations. 
Louvain/Leiden is run a large number of times for each $t$ starting from different random initializations to obtain an ensemble of optimized solutions. 

To evaluate the consistency of this ensemble of solutions, we use the normalized variation of information (NVI)~\cite{kraskovHierarchicalClusteringBased2003}, which measures the distance between partitions. 
We thus compute the average NVI($t$)
between all pairs of partitions (or a random subset thereof to reduce computational cost) obtained at scale $t$. 
A low value of the average NVI$(t)$ indicates a reproducible (robust) solution for the optimization~\eqref{eqn:gen_mod}, suggesting a well-defined maximum and hence increased confidence in the optimal partition found. 

This quantitative notion of robustness is also applied to compare the partitions obtained across scales by computing NVI$(t,t')$, i.e., the distance between the optimal partitions at scales $t$ and $t'$. In this case, persistently low NVI$(t,t')$ across a long stretch of $t$ indicates that a partition (or a set of similar partitions) is found robustly across graph scales.

\paragraph{Post-processing of optimal partitions.}
Given the greedy nature of the Louvain/Leiden optimizers, it is
possible that the optimal partition found at scale $t'$ could in fact be a better partition for scale $t$ than the partition found by Louvain/Leiden at  $t$. We run a post-processing step that checks for and selects any such improved partition for scale $t$ even if found at any other $t'$.

\paragraph{Automated scale selection.}
We aim to find relevant scales at which partitions are robust both with respect to the optimization (low NVI$(t)$) and across scales (extended blocks of low NVI$(t,t')$). The partitions found at such scales give a good description of the graph structure at a level of coarseness (or resolution). 
The selection of scales can be done by visual inspection of the result summary plot, see Fig.~\ref{fig:example}, or using the automated scale selection criterion introduced by~\cite{schindlerMultiscaleMobilityPatterns2023}, which combines the robustness to the optimization and the persistence across scales. 

\subsection{Main parameters and default values.}

To make \texttt{PyGenStability} easier to use for non-experts, we have set default values for several parameters (default values in parentheses below).
The chosen quality function is optimised over \texttt{n\_scale} (= 20) scales, chosen equidistantly between \texttt{min\_scale} (= -2.0) and \texttt{max\_scale} (= 0.5) on a log scale. Hence \texttt{min\_scale} and \texttt{max\_scale} determine the minimal and maximal coarseness of the partitions, respectively, and \texttt{n\_scale} increases the resolution of the analysis. 
Operationally, we recommend starting with the default \texttt{n\_scale} and increasing it for more fine-grained results.

To quantify the robustness of the partitions with respect to the optimisation of the quality function, an ensemble of \texttt{n\_tries} (= 100) solutions is computed using Louvain or Leiden, and the similarity of the solutions is estimated by computing the average pairwise NVI$(t)$ of a random subset of \texttt{n\_NVI} (= 20) partitions. Increasing \texttt{n\_tries} leads to a better estimation of the robustness, at a computational cost since the total number of Louvain/Leiden optimizations performed is \texttt{n\_scale} $\times$ \texttt{n\_tries}. 

The scale selection follows a sequential algorithm developed in Ref.~\cite{schindlerMultiscaleMobilityPatterns2023}. 
To detect intervals over which partitions remain similar, we apply average pooling to NVI$(t,t')$ with kernel size, \texttt{kernel\_size} (= 0.1 $\times$ \texttt{n\_scale}), followed by smoothing of its diagonal with a triangular moving mean, where the smoothness is controlled by the window size, \texttt{window\_size} (= 0.1 $\times$ \texttt{n\_scale}).  This gives the curve  \textit{Block NVI}$(t)$. Increasing \texttt{kernel\_size} enlarges the interval over which scales need to be persistent and increasing \texttt{window\_size} further smoothes out random variability across scales.
We then define basins with radius \texttt{basin\_radius} (= 0.01 $\times$ \texttt{n\_scale}) around all the local minima of \textit{Block NVI}$(t)$.
From each basin, we select a scale, given by the solution with minimal NVI$(t)$ within the basin.
This procedure selects scales that are both persistent across $t$ and robust to the combinatorial optimisation.

\section{Benchmarking}

\begin{figure}[htpb]
    \centering
    \includegraphics[width=\textwidth]{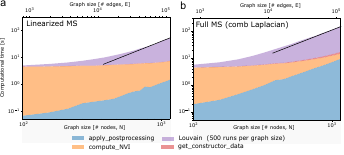}
    \caption{{\bf Code benchmarking.} To assess the computational efficiency and scalability of each component of the code, we analysed stochastic block model (SBM) graphs of increasing size. (An example of these graphs with $N=270$ nodes is shown in Fig.~\ref{fig:example}.) We show benchmarking results for {\bf (a)} `modularity with resolution parameter', equivalent to linearized Markov Stability (MS), and {\bf (b)} MS with combinatorial graph Laplacian. The computations were performed on a single CPU and involved $500$ Louvain optimizations for each graph size (i.e., $50$ Louvain runs computed at $10$ scales $t$). The cost of the 'Louvain optimization' and 'post-processing' steps increase with graph size more sharply for full MS in (b) as compared to linearised MS in (a). This is due to the decreased sparsity of the quality matrix, and the computational cost scales approximately as $O(E)$ (black line), where $E$ is the number of edges of the graph. 
    }
    \label{fig:benchmark}
\end{figure}

To assess computational efficiency as a function of graph size, we timed the core functions called during a computation with Louvain for: (i) linearized MS, and (ii) MS with combinatorial graph Laplacian (Fig.~\ref{fig:benchmark}).
We find that the rate-limiting function for larger graphs is the Louvain optimization, followed by the growing computational cost of obtaining the matrix exponential, whereas the other computations have a near-constant computational cost. Hence, for large graphs, we provide the linearized MS quality function to avoid the loss of sparsity induced by the matrix exponential. 
\section{Example and applications}

\begin{figure}[htb!]
   \centering
   \includegraphics[width=\textwidth]{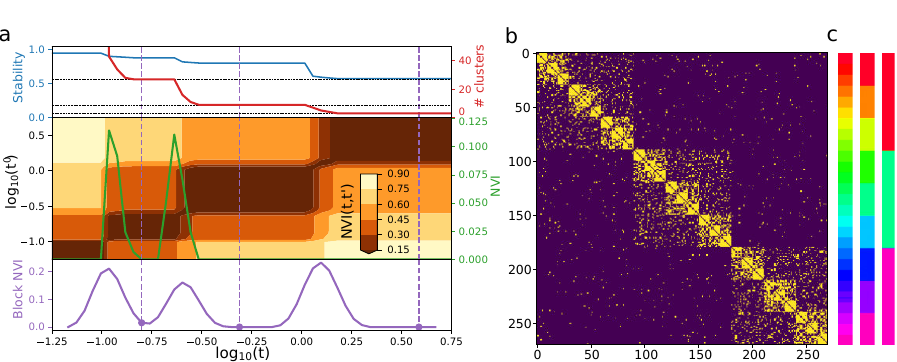}
   \caption{{\bf Example of multiscale community detection. }
   {\bf (a)} Summary plot for a multiscale SBM graph. The top row shows the value of the optimized Generalized MS function $Q_{gen}(H^*(t))$ together with the number of communities in the optimal partition $H^*(t)$ as a function of the scale $t$. The second row shows the two robustness measures for the obtained partitions: NVI$(t)$ for each scale and NVI$(t, t')$ across scales. The bottom row shows the automated scale selection criterion, with basins corresponding to blocks in NVI$(t, t')$ and robust scales identified as local minima of NVI$(t)$ within each basin (purple dots). 
   {\bf (b) } Adjacency matrix of the graph in this toy example: a multiscale SBM graph with $N=270$ nodes and ground truth of 3 planted scales with 27, 9, and 3 clusters.
   {\bf (c)} The communities determined by the scale selection criterion in (a) are indicated by different colours for the three detected scales and correspond to the ground truth.
   }
   \label{fig:example}
\end{figure}

As a simple illustration of the use of the package, we provide an example of the multiscale analysis of a toy graph: a multi-scale SBM with planted partitions at three scales. Fig.~\ref{fig:example} shows that \texttt{PyGenStability} is able to accurately recover the expected partitions at the three scales. 

The MS framework, which is now made available through \texttt{PyGenStability}, has already been used extensively to analyse multiscale community structures in real-world graphs, also called \textit{networks} in the literature, from diverse domains facilitating a range of applications. These include detecting functional and anatomical constituents in the directed neuronal network of \textit{C. elegans}~\cite{bacikFlowBasedNetworkAnalysis2016}, interest communities in the Twitter network of the 2011 UK riots~\cite{beguerisse-diazInterestCommunitiesFlow2014}, spatial and dynamical subunits in protein structures~\cite{delmotte2011protein, peach2019unsupervised}, hospital catchment areas in surgical admission networks~\cite{clarkeDefiningHospitalCatchment2019}, learning behaviours among online students~\cite{peach2019data}, multiscale human mobility patterns under lockdown~\cite{schindlerMultiscaleMobilityPatterns2023} and in hospitals~\cite{myall2021network} during COVID-19, topic modelling with semantic networks derived from free text~\cite{altuncuFreeTextClusters2019}, and quantifying information flow and bottlenecks using discrete network geometry~\cite{Gosztolai2021}. Detailed illustrations and examples of applications to several synthetic and real-world networks are provided as examples in the code, including an analysis of a power grid network and protein structural graphs. 

\section{Conclusion and outlook}

The Python package \texttt{PyGenStability} is primarily designed for multiscale community detection within the MS framework but can be extended for the optimization of a range of graph clustering quality functions. \texttt{PyGenStability} allows researchers to identify robust graph partitions at different resolutions in graphs of different types and has been already applied widely to unsupervised learning tasks for real-world networks from various domains. In future work, we plan to further improve the automatic scale selection functionality, extend the range of constructors for different quality functions, and perform a quantitative comparison of the multiscale optimization using the Louvain and Leiden algorithms.
    
\section*{Acknowledgements}
We thank Vincent Traag for help with the implementation of the Leiden optimizer in PyGenStability.
Funding in direct support of this work: 
AA, RP and MB acknowledge funding through the EPSRC award EP/N014529/1 supporting the EPSRC Centre for Mathematics of Precision Healthcare at Imperial.
RP acknowledges funding from the Deutsche Forschungsgemeinschaft (DFG, German Research Foundation) Project-ID 424778381-TRR 295.
JS acknowledges support from the EPSRC (PhD studentship through the Department of Mathematics at Imperial College London). 
AG acknowledges support from an HFSP Cross-disciplinary Postdoctoral Fellowship (LT000669/2020-C).

\setlength\bibitemsep{0pt}

\printbibliography

\end{document}